\documentclass[longauth]{aa} 
\usepackage{graphicx}
\usepackage{subfig}
\usepackage[round]{natbib}
\usepackage{color}
\usepackage{amsmath}
\usepackage[normalem]{ulem}

\newcommand{\cmc}{\mbox{$\mbox{cm}^{-3}$}}
\newcommand{\cmd}{\mbox{$\mbox{cm}^{-2}$}}

\newcommand{\cmg}{\mbox{$\mbox{cm}^{2} \, \mbox{g}^{-1}$}} 

\newcommand{\Msun}{\mbox{$M_\odot$}}
\newcommand{\msun}{\mbox{$M_\odot$}}

\newcommand{\Nhtwo}{$N_{\rm H_2}$}
\newcommand{\nhtwo}{\mbox{$n_{\rm H_2}$}\,}

\newcommand{\msmmT}{\mbox{$M_{\mbox{\tiny 3.4\,mm}}^{\rm 20\,K, 6\,kpc}$}}
\newcommand{\tdust}{\mbox{$T_{\tiny{\rm dust}}$}}
\newcommand{\kmm}{\mbox{$\kappa_{\mbox{\tiny 3.4\,mm}}$}}
\newcommand{\sint}{\mbox{$S_{\mbox{\tiny 3.4\,mm}}^{\mbox{\tiny ~int}}$}}
\newcommand{\Bmm}{\mbox{$B_{\mbox{\tiny 3.4\,mm}}$(\tdust)}}

\definecolor{gris}{gray}{0.5}

\begin{document}

\title{The W43-MM1 mini-starburst ridge, a test for star formation efficiency models}
  \author{F. Louvet\inst{1}
  \and F. Motte\inst{1}
  \and P. Hennebelle \inst{1}
  \and A. Maury\inst{2,1}
  \and I. Bonnell \inst{3}    
  \and S. Bontemps\inst{4}
  \and A. Gusdorf\inst{5}
  \and T. Hill\inst{6}
  \and F. Gueth\inst{7}
  \and N. Peretto \inst{8}
  \and A. Duarte-Cabral \inst{9}
  \and G. Stephan\inst{10,}\inst{11}
  \and P. Schilke\inst{10}
  \and T. Csengeri\inst{12}
    \and Q.~Nguy$\tilde{\hat{\rm e}}$n~Lu{\hskip-0.65mm\small'{}\hskip-0.5mm}o{\hskip-0.65mm\small'{}\hskip-0.5mm}ng\inst{13}
  \and D. C. Lis\inst{14,}\inst{15}
          }
\institute{Laboratoire AIM Paris-Saclay, CEA/IRFU - CNRS/INSU - Universit\'e Paris Diderot, Service d'Astrophysique, B\^at. 709, CEA-Saclay, F-91191 Gif-sur-Yvette Cedex, France, 
              \email{fabien.louvet@cea.fr}
  \and Harvard-Smithsonian Center for Astrophysics, Cambridge, MA
  \and Scottish Universities Physics Alliance (SUPA), School of Physics and Astronomy, University of St. Andrews, North Haugh, St Andrews, Fife KY16 9SS, UK
  \and OASU/LAB-UMR~5804, CNRS/INSU - Universit\'e Bordeaux 1, 2 rue de l'Observatoire, BP 89, F-33270 Floirac, France 
  \and LERMA, UMR 8112 du CNRS, Observatoire de Paris, \'Ecole Normale Sup\'erieure, 24 rue Lhomond, 75231 Paris Cedex 05, France
  \and { Joint ALMA Observatory, Alonso de Cordova 3107, Vitacura, Santiago, Chile}
  \and  Institut de Radioastronomie Millim\'etrique (IRAM), 300 rue de la Piscine, 38406 Saint Martin d'H\`eres, France
  \and School of Physics and Astronomy, Cardiff University, Queens Buildings, The Parade, Cardiff CF24 3AA, UK
  \and School of Physics and Astronomy, University of Exeter, Stocker Road, Exeter EX4 4QL, UK
  \and Physikalisches Institut, Universit\"at zu K\"oln, Z\"ulpicher Str. 77, D-50937 K\"oln, Germany
  \and LERMA2, Observatoire de Paris, 61 Av. de l'Observatoire, 75014 Paris, France
  \and Max-Planck-Institut f\"ur Radioastronomie, Auf dem H\"ugel 69, D-53121 Bonn, Germany           
   \and Canadian Institute for Theoretical Astrophysics, University of Toronto, 60 St. George Street, Toronto, ON M5S~3H8, Canada 
  \and California Institute of Technology, Pasadena, CA 91125, USA
  \and Sorbonne Universit\'{e}s, Universit\'{e} Pierre et Marie Curie, Paris 6, CNRS, Observatoire de Paris, UMR 8112, LERMA, Paris, France
        }

     \date{Received 2014; accepted 2014}


  \abstract
  {Star formation efficiency (SFE) theories are currently based on statistical distributions of turbulent cloud structures and a simple model of star formation from cores. They remain poorly tested, especially at the highest densities.}
  %
   {We investigate the effects of gas density on the SFE through measurements of the core formation efficiency (CFE). With a total mass of $\sim$2$\times 10^4~\msun$, the W43-MM1 ridge is one of the most convincing candidate precursor of Galactic starburst clusters and thus one of the best places to investigate star formation.}
  %
   {We used high-angular resolution maps obtained at 3~mm and 1~mm  within the W43-MM1 ridge with the IRAM Plateau de Bure Interferometer to reveal a cluster of 11 massive dense cores (MDCs), and, one of the most massive protostellar cores known. An \emph{Herschel} column density image provided the mass distribution of the cloud gas. 
   We then measured the \lq instantaneous\rq~CFE and estimated the SFE and the star formation rate (SFR) within subregions of the W43-MM1 ridge.}
  %
   {The high SFE found in the ridge ($\sim$6\% enclosed in $\sim$8 pc$^3$) confirms its ability to form a starburst cluster. There is however a clear lack of dense cores in the northern part of the ridge, which may be currently assembling.
   The CFE and the SFE are observed to increase with volume gas density while the SFR per free fall time steeply decreases with the virial parameter, $\alpha_{vir}$. Statistical models of the SFR may well describe the outskirts of the W43-MM1 ridge but struggle to reproduce its inner part, which corresponds to measurements at low $\alpha_{\rm vir}$. It may be that ridges do not follow the log-normal density distribution, Larson relations, and stationary conditions forced in the statistical SFR models.}
{}

\titlerunning{Fragmentation of the W43-MM1 mini-starburst ridge}
\authorrunning{Louvet et al.}
   \maketitle
%

\section{Introduction}
The formation of high-mass stars remains poorly understood but an emerging scenario suggests that they form in massive dense cores (MDCs: $\sim$0.1~pc and $>10^5~\cmc$ as defined in \citealt{motte07}, see also \citealt{wang14}) through dynamical processes such as colliding flows initiated by cloud formation \citep[e.g.][]{csengeri11a,quang13}. 
The \emph{Herschel} key program HOBYS \citep[see][]{motte10hobys,motte12hobys} identifies ridges as high-density filaments, above $10^{23}$~cm$^{-2}$ in column density, favorable to the formation of high-mass (OB-type, $\geq$8~\msun) stars \citep[see][]{hill11-vela,quang11-w48,martin12}. The most extreme of these ridges, W43-MM1, lies in the massive, highly concentrated and very dynamic W43 molecular complex located at 6 kpc \citep{quang11-w43,carlhoff13}. In its central region, W43-MM1 is thought to be experiencing a cloud collision (\citealt{quang13}), causing a remarkably efficient burst of high-mass star formation \citep{motte03}.
The W43-MM1 ridge can be modeled by a 3.9~pc$\,\times\,2$~pc$\,\times\,$2~pc ellipsoid with a total mass of $\sim$2$\times 10^4$~M$_\odot$ and an average density of $\sim$4.3$\times10^4~\rm cm^{-3}$, physically large enough and massive enough to form a large cluster. 
Its fragmentation has been studied before, with a 0.2 pc resolution, by \cite{motte03}. Fragmentation,
 magnetic field, outflows, and hot core of the densest part of the W43-MM1 ridge has also been observed with high-angular resolution by \cite{cortes06} and \cite{sridharan14}.

A handful of studies have been carried out to estimate the core formation efficiency (CFE) in high-mass star forming regions, and suggested that the stellar formation efficiency (SFE) increases with gas density (\citealt{bontemps10}, \citealt{palau13}). As for the stellar formation rates (SFRs), most statistical models directly relate it to the amount of gas above a given density threshold \citep[][]{krumholz05,padoan11,hennebelle11}. If this view agrees with the SFR measurements in low-mass star-forming clouds \citep{heiderman10}, which are found to be proportional to cloud masses \citep[Eq.~3 of][]{lada10,evans14}, they are not representative of typical Galactic clouds forming high-mass stars (\citealt{motte03}, \citealt{quang11-w48}). These observational differences cast doubt on the accuracy of extrapolating scaling laws observed in low-mass star-forming regions to describe star formation in clouds forming high-mass stars.




MDCs hosting high-mass protostars can be used to investigate the fragmentation of ridges and measure the concentration of its gas into high-density seeds and then high-mass stars. 

In the present paper\footnote{
Based on observations carried out with the IRAM Plateau de Bure Interferometer. IRAM is supported by INSU/CNRS (France), MPG (Germany) and IGN (Spain).
}, we investigate the CFE variations through the W43-MM1 ridge and compare the resulting SFE \& SFR estimates to predictions of star formation models. Section~\ref{s:obs} presents an interferometric imaging of W43-MM1 that reveals a cluster of MDCs characterized in Sect.~\ref{s:census}. Section~\ref{s:cfe} presents an analysis of the CFE in subregions of the ridge and discusses the CFE variations with cloud volume density. In Section~\ref{s:sfe-sfr} we present two methods to compute the SFEs and the SFRs from the observed CFEs in W43-MM1. Finally, the SFR measured in the different subregions of the ridge are compared to predictions of statistical models of star formation in Sect.~\ref{s:sfrff}.

\section{Observations, reduction and dataset}
\label{s:obs}

\subsection{IRAM/PdBI}

A seven-field 3~mm mosaic of the W43-MM1 ridge and a single 1~mm pointing toward W43-N1, its most massive dense core, have been carried out with the IRAM Plateau de Bure Interferometer (hereafter IRAM/PdBI, see Table~\ref{t:parpdbi}). 
Configurations C2 \& D with 4-6 antennas were used in March-April \& October-November 2002 for the single pointing toward the phase center $(\alpha,\delta)=$ 18:47:47.1,    -01:54:28; configurations C \& D were used in October \& July 2011 with respectively 5 and 6 antennas for the mosaic.
Broad-band continuum and spectral lines (not shown here) were simultaneously observed.
The phase, amplitude, and correlator bandpass were calibrated on strong quasars (3C273, 4C09.57 and 1936-155 in 2002; 3C454.3, 1827+062, and 0215+015 in 2011) while the absolute flux density scale was derived from MWC349 observations. The absolute flux calibration uncertainty is estimated to be $\sim$15\%.

The two WIDEX subunits were combined to observe the continuum emission with a total bandwidth of 3.6 GHz centered at 87.5 GHz (3~mm). Two correlator units were summed into a 640~MHz bandwidth centered at 239.5~GHz (1~mm). The mean angular resolutions were respectively $3\farcs96$ at 3~mm and $2\farcs19$ at 1~mm.

\begin{figure*}[htbp!]
\begin{center}
\includegraphics[trim=0cm 0cm 0cm 0cm, clip, scale=0.6]{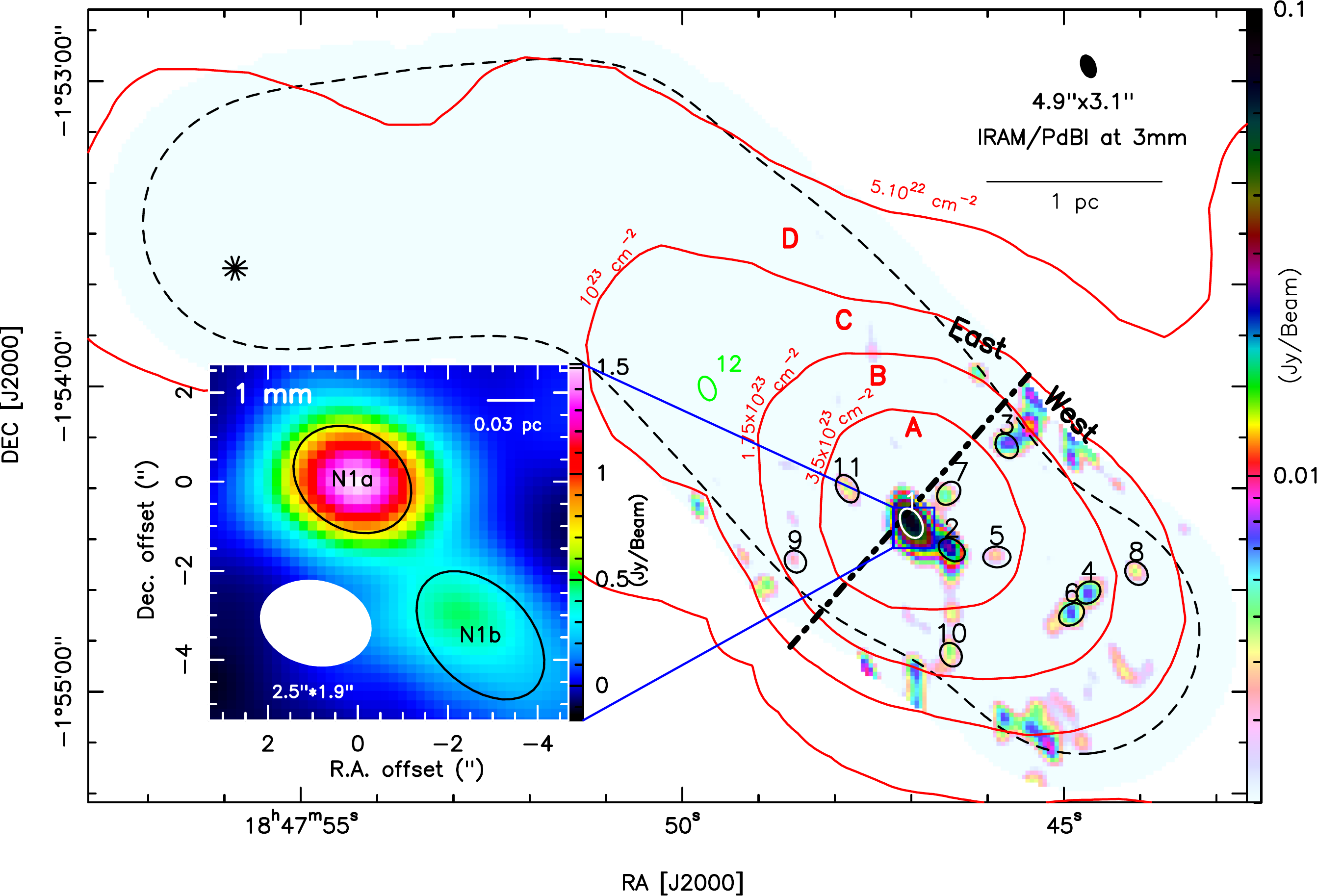}
\caption{IRAM/PdBI 3~mm continuum image of the W43-MM1 ridge revealing a cluster of MDCs. The black dashed contour outlines the area where the confidence weights map exceeds 10\%. Column densities yielded by \emph{Herschel}  imaging \citep[see][]{quang13} and shown in red contours are used to define A--D subregions. The black dotted-dashed line and the 10$^{23}$ cm$^{-2}$ contour outline the eastern and western parts of the ridge. Black and white ellipses plus numbers locate MDCs extracted by \emph{Getsources} (see Table \ref{t:tableobj}), the green ellipse outlines N12, and the black asterisk pinpoints a source identified by \citet{beuther12w43}. Note that negative contours have been removed to reduce confusion. See Appendix for the figure with negative contours.
{\bf Zoom inset:}
IRAM/PdBI 1~mm continuum image of the W43-N1 MDC. Black ellipses are HMPCs extracted by \emph{Getsources} (see Table~\ref{t:tableobj}).
}
\label{oncont}
\end{center}
\end{figure*}

We used the GILDAS\footnote{
The Grenoble Image and Line Data Analysis Software is developed and maintained by IRAM to reduce and analyze data obtained with the 30~m telescope and Plateau de Bure interferometer. See
www.iram.fr/IRAMFR/GILDAS
} package to calibrate each dataset, merge the visibility data of all fields for the 3~mm mosaic, then invert and clean (natural cleaning) both the 1~mm and 3~mm.
We built a `pure' continuum map at 3~mm from spectral bands of WIDEX free of strong lines. 
The 3~mm continuum map is given in Fig.~\ref{oncont}. It displays a very inhomogeneous repartition of the continuum with much more emission in the south-western part of the map. Due to limited dynamic range around the strong continuum and extended source W43-N1, we obtain significantly different rms levels in the north-eastern part of the 3~mm mosaic (3$\sigma\sim$ 0.11 mJy/beam) than in the southern region around W43-N1 (3$\sigma\sim$ 3.8 mJy/beam). As for the 1 mm pointing, we measured a 3$\sigma$ rms level of $\sim$ 0.15 Jy/beam.


\begin{table*}[htbp!]
\begin{center}
\addtolength{\tabcolsep}{-0pt}
\caption{Main observational parameters}
\label{t:parpdbi}
\begin{tabular}{lccc}
\hline
\hline
Parameter                  & 3~mm                        & 1~mm                           & \emph{Herschel} column density        \\
\hline
Frequency                  & 87.43$^{a}$ GHz             & 239.5 GHz                      & --                                    \\
Bandwidth                  & 3600 MHz                    & 640 MHz                        & --                                    \\
System temperature         & $\sim$120 K                 & $\sim$300 K                    & --                                    \\
Primary beam               & 59\arcsec                   & 21\arcsec                      & --                                    \\
Synthesized beam           & $4\farcs85\times3\farcs06$  & $2\farcs51\times1\farcs92$     & 25\arcsec                             \\
$3\sigma$ rms              &$0.11-3.8$~mJy/beam          & $\sim$150~mJy/beam             & $\sim4.5\times 10^{21}$cm$^{-2}$      \\
\hline
\end{tabular}
\end{center}
Note: $^a$ The mean frequency was calculated assuming a S($\nu$)$\propto \nu^{-2}$ emission spectrum accurately describing the ISM SED slope in the WIDEX band. 
\end{table*}

\subsection{Herschel dust temperature and column density maps}

We used the dust temperature and column density images built from Hi-GAL and HOBYS data \citep{molinari10,motte10hobys} and presented by \cite{quang13}. Using three of the four longest wavelengths of \emph{Herschel} (160–-350 $\mu$m), they derived the total (gas+dust) column density ($N_{\rm H_2}$) and average dust temperature maps of W43-Main with an angular resolution of 25$\arcsec$ (see Table~\ref{t:parpdbi}). Following the procedure fully described in \citet{hill11-vela,hill12-vela}, they fitted pixel-by-pixel spectral energy distributions (SEDs) with modified blackbody models. They used a dust opacity law similar to that of \cite{hildebrand83} but with $\beta = 2 $ instead of $\beta = 1 $ and assumed a gas-to-dust ratio of 100: $\kappa_{\nu} = 0.1 \times (300 \mu \rm{m}/\lambda)^2$ cm$^2$ g$^{-1} $. It provides column density images with very low (20\% when A$_v>$ 10 mag) relative uncertainties, arising from SED fit errors \citep{hill09} and possible variations of the emissivity index through the map. The absolute accuracy of \emph{Herschel} $N_{\rm H_2}$  maps has been estimated to be around 40\% \citep{roy14}.

\section{MDCs census in the W43-MM1 ridge}
\label{s:census}


\begin{table*}[htbp!]
\begin{center}
\addtolength{\tabcolsep}{-1.8pt}
{
\caption{MDC and HMPC samples}
\label{t:tableobj}
\begin{tabular}{lccccccccl}
\hline
\hline
Name &  R.A. & Dec. & \it Size & $d$& $S^{\rm int}$& \emph{FWHM}$^a$ & 
$M^{\rm 20\,K}~^b$  & $<\nhtwo>^c$  & Remarks   \\
& (J2000)&(J2000)&[$\arcsec\times\arcsec$] &[kpc]&[mJy]&[pc] & [\msun] & [10$^7$\cmc] &\\
&(1)&(2)&(3)&(4)&(5)&(6)&(7)&(8) \\
\hline
N1 MDC            & 18:47:47.0 & -1:54:26   & $6.2\times4.0$  &6   & 135.2      & 0.088   &  2128$\pm$53   & $10.35\pm0.26$    & \\
N2 MDC            & 18:47:46.4 & -1:54:32   & $5.6\times4.0$  &6   & 34.7       & 0.076   &  545$\pm$59    & $4.19\pm0.45$	    & \\
N3 MDC            & 18:47:45.7 & -1:54:11   & $5.2\times4.0$  &6   & 13.4       & 0.066   &  211$\pm$38    & $2.43\pm0.47$     & \\
N4 MDC            & 18:47:44.6 & -1:54:40   & $5.0\times4.1$  &6   & 9.7        & 0.074   &  153$\pm$38    & $1.92\pm0.47$     & \\
N5 MDC            & 18:47:45.8 & -1:54:33   & $5.4\times4.1$  &6   & 9.7        & 0.074   &  153$\pm$27    & $1.24\pm0.22$     & \\
N6 MDC            & 18:47:44.9 & -1:54:44   & $5.3\times4.0$  &6   & 8.9        & 0.069   &  141$\pm$36    & $1.45\pm0.37$     & \\
N7 MDC            & 18:47:46.5 & -1:54:21   & $5.1\times4.0$  &6   & 7.1        & 0.063   &  111$\pm$26    & $1.45\pm0.33$     & \\
N8 MDC            & 18:47:44.0 & -1:54:36   & $5.0\times4.0$  &6   & 6.7        & 0.061   &  105$\pm$19    & $1.56\pm0.28$     & \\
N9 MDC            & 18:47:48.5 & -1:54:34   & $4.6\times4.0$  &6   & 3.5        & 0.048   &  56$\pm$17     & $1.64\pm0.49$     & \\
N10 MDC           & 18:47:46.4 & -1:54:52   & $4.9\times4.0$  &6   & 5.6        & 0.059   &  88$\pm$26     & $1.50\pm0.45$     & \\
N11 MDC           & 18:47:47.8 & -1:54:20   & $5.8\times4.0$  &6   & 7.3        & 0.080   &  115$\pm$24    & $0.75\pm0.16$     & \\
\hline                                                                                                                                                        
N12 dense core    & 18:47:49.6 & -1:54:00   & $5.2\times4.1$  &6   & 1.3        & --      &  $21\pm5$      & --                & aperture extraction   \\
N1a HMPC          & 18:47:47.0 & -1:54:26   & $2.8\times2.2$  &6   & 1900       & 0.033   &  1080$\pm$ 35  & $104\pm3.4$       & 1~mm extraction       \\
N1b HMPC          & 18:47:46.8 & -1:54:29   & $3.4\times2.2$  &6   & 700        & 0.047   &  395$\pm$30    & $12.9\pm1$        & 1~mm extraction       \\
\hline
\multicolumn{3}{l}{Cygnus~X MDCs}       &    & 1.4$^d$         & -       &$\sim$0.11$^d$    &$\sim$60$^{d,e}$      &$\sim$0.18$^{d,e}$        & Motte et al. 2007     \\
\multicolumn{3}{l}{SDC335-MM1 HMPC}     &    & 3.25            & -       &$\sim$0.054       &$\sim$343$^e$         &$\sim$7$^e$               & Peretto et al. 2013   \\
\multicolumn{3}{l}{G0.22\&G0.24 HMPCs}  &    & 3.6--6.5        & -       &$\sim$0.034       &$\sim$8.7$^e$         &$\sim$1$^e$               & Rathborne et al. 2007 \\
\multicolumn{3}{l}{G11.11–0.12 HMPCs}   &    & 3.6             & -       &$\sim$0.025       &$\sim$10.6$^e$        &$\sim$2.94$^e$            & Wang et al. 2014      \\
\multicolumn{3}{l}{Cygnus~X HMPCs}      &    & 1.4$^d$         & -       &$\sim$0.016$^d$   &$\sim$12$^{d,e}$      &$\sim$3.6$^{d,e}$         & Bontemps et al. 2010  \\
\hline	
\end{tabular}}
\end{center}
~$^a$ \emph{FWHM}s are sizes of Col.~3 deconvolved by the beam and set at 6 kpc distance:
\emph{FWHM}$=\sqrt{{\mbox{\it Size}_{\rm\tiny major}}\times{\mbox{\it Size}_{\rm\tiny minor}}-\mbox{\it HPBW}^2}\times d$.\\
~$^b$ $M^{\rm 20\,K}$ masses and relative uncertainties are calculated with Eq.~\ref{pedro} from integrated fluxes $S^{\rm int}$ (Col. 4) and errors measured by \emph{Getsources}, except when mentioned in remarks. The errors do not take into account the absolute uncertainties such as flux calibration, temperature and emissivity assumptions. \\
~$^c$ Mean densities are measured from Cols.~4 and 5 via $<\nhtwo> = \frac{M^{\rm 20\,K}}{\frac{4}{3}\;\pi\;\times\;(\mbox{\it \small FWHM}/2)^{\;3}}$.\\
~$^d$ Sizes, masses, and densities have been recalculated with a distance to Cygnus X of 1.4 kpc from the Sun \citep{rygl12}. \\
~$^e$ Measurements have been re-calculated using Eq.~1, or its equivalent at 1~mm, T=20 K, dust opacity discussed in Sect. 3, and the equation given in c. \\
\end{table*}


To extract the MDCs, we used the source extraction tool \emph{Getsources} \citep{sacha12}. Developed for multi-wavelength \emph{Herschel} images, it calculates the local noise and local background to properly extract compact sources from a complex cloud environment. We increased the quality constraints\footnote{
Input parameters `sreliable' and `cleantuning' were multiplied by two with respect to the default values of Getsources. The parameter `sreliable' controls the significance of reliable sources in the extraction catalogs and the parameter `cleantuning' adjusts the cleaning depth.}
of \emph{Getsources} to account for the specificity of our interferometric images. To perform a confident extraction of MDCs, we masked the map borders where confidence map weights drop below 10\% (dashed contour in Fig.~\ref{oncont}). We also set a maximum source size of 0.25~pc to focus on 0.1~pc MDCs at 3~mm, and, 0.02~pc at 1~mm to focus on 0.01~pc high-mass protostellar cores (HMPCs).

At 3~mm, \emph{Getsources} identified 11 MDCs, with average deconvolved sizes of $\sim$0.07 pc, all located in the densest south-western part of the
 W43-MM1 ridge (see Fig.~\ref{oncont} and Table~\ref{t:tableobj}). 
 Four MDCs are substructures of $\sim0.2$ pc clumps extracted by \cite{motte03} 
 and four correspond to the $\sim0.02$ pc HMPCs identified by \cite{sridharan14}. 
 Among our MDCs, six have outflows
 (Louvet et al. in prep.). Only N12, the less massive of our sample ($\sim$20 M$_\odot$), was suggested by outflows but not extracted by \emph{Getsources}. \citet{beuther12w43} detected another diffuse dust source of $\sim$30\arcsec~size which remains undetected (see asterisk in Fig.~\ref{oncont}), likely filtered out by the interferometer (filtering scale $\sim$20\arcsec).
The map shown in Fig.~\ref{oncont} suggests an uneven distribution of the dense gas, with 3 MDCs forming in the eastern part of the ridge and 8 MDCs in its western part.

The 1~mm map only covers the W43-N1 MDC (see Fig.~\ref{oncont}). It shows that this core splits into two HMPCs (see Fig.~\ref{oncont} and Table~\ref{t:tableobj}) with sizes approaching that of protostellar envelope scales \citep[e.g.][]{rathborne07,bontemps10}.

To derive the masses of the MDCs in our sample, we assumed that the 3~mm continuum emission mainly arises from thermal dust and is optically thin. The free-free contribution to the 3~mm fluxes is estimated to be much less than 20\% for the MDCs. Indeed the noise peaks found in the Cornish survey at 2~cm \citep{hoare12} and extrapolated at 3.4~mm assuming an optically thin free-free emission spectral index correspond to a free-free contamination of $\sim$0\% for most MDCs up to $\sim$20\% for N7.

The MDC masses are calculated from the integrated fluxes measured by \emph{Getsources}, $\sint$, via:
 \begin{equation*}
M_{\rm 3.4 mm} =\frac{\sint \times d^2}{\kappa_{\rm 3.4 mm}}\times\frac{1}{B_{\rm 3.4 mm}}~,
 \end{equation*}
with $d$ the distance from the Sun and $\Bmm$ the Planck function. The dust mass opacity was taken equal to $\kmm$~=~$2.6\times 10^{-3}~\cmg$, following the $\kappa_\nu =0.1~\cmg \times (\nu/1000~\mbox{GHz})^\beta$ equation with an opacity index $\beta=1.5$ which is typical for dense and cool media \citep{ossenkopf94}. 

We used $\tdust=20$~K as suggested by the averaged \emph{Herschel} dust temperature map over the W43-MM1 ridge (see Sect.~\ref{s:obs}) and the dust temperature estimated for the W43-MM1 clump only \citep{motte03,bally10}.
The formula results in:
\begin{equation}
\msmmT =158~\Msun\,\times\frac{\sint}{\rm 0.01~Jy}
\times\frac{\kmm}{\rm 2.6\times10^{-3}~\cmg}  
\label{pedro}
\end{equation}

Errors on the $\msmmT$ masses mostly arise from the dust mass opacity at 3.4~mm, which could be a factor 3.5 smaller, if one uses an optical index of $\beta$=2, thus increasing the masses by a similar factor. Moreover, since N1 hosts a hot core \citep[e.g. ][]{herpin12,sridharan14}, its temperature could be higher than 20 K. The temperature constraints presented in \cite{motte03} and \cite{sridharan14}, namely 20 K and 300 K for a FWHM of 0.25 pc and 0.017 pc respectively, suggest a temperature profile of $T\propto FWHM^{-1}$. This would leads to a temperature of 55 K for the N1 MDC, and would decrease its mass down by a factor of 3. 

The masses of HMPCs at 1.3~mm (see Table~\ref{t:tableobj}) were calculated from an equivalent equation to Eq.~\ref{pedro} at 1.3~mm, using $\kappa_{\rm 1.3mm}= 0.01~\cmg$. The dust opacity uncertainties could contribute on errors on the mass measurements up to a factor 2.
The application of the temperature profile derived above ($T\propto FWHM^{-1}$) would decrease the 1.3~mm mass of N1a by a factor of 10. 

\begin{figure*}[htbp!]
 \begin{center}
\includegraphics[trim=0cm 0cm 0cm 0cm, clip, scale=0.5]{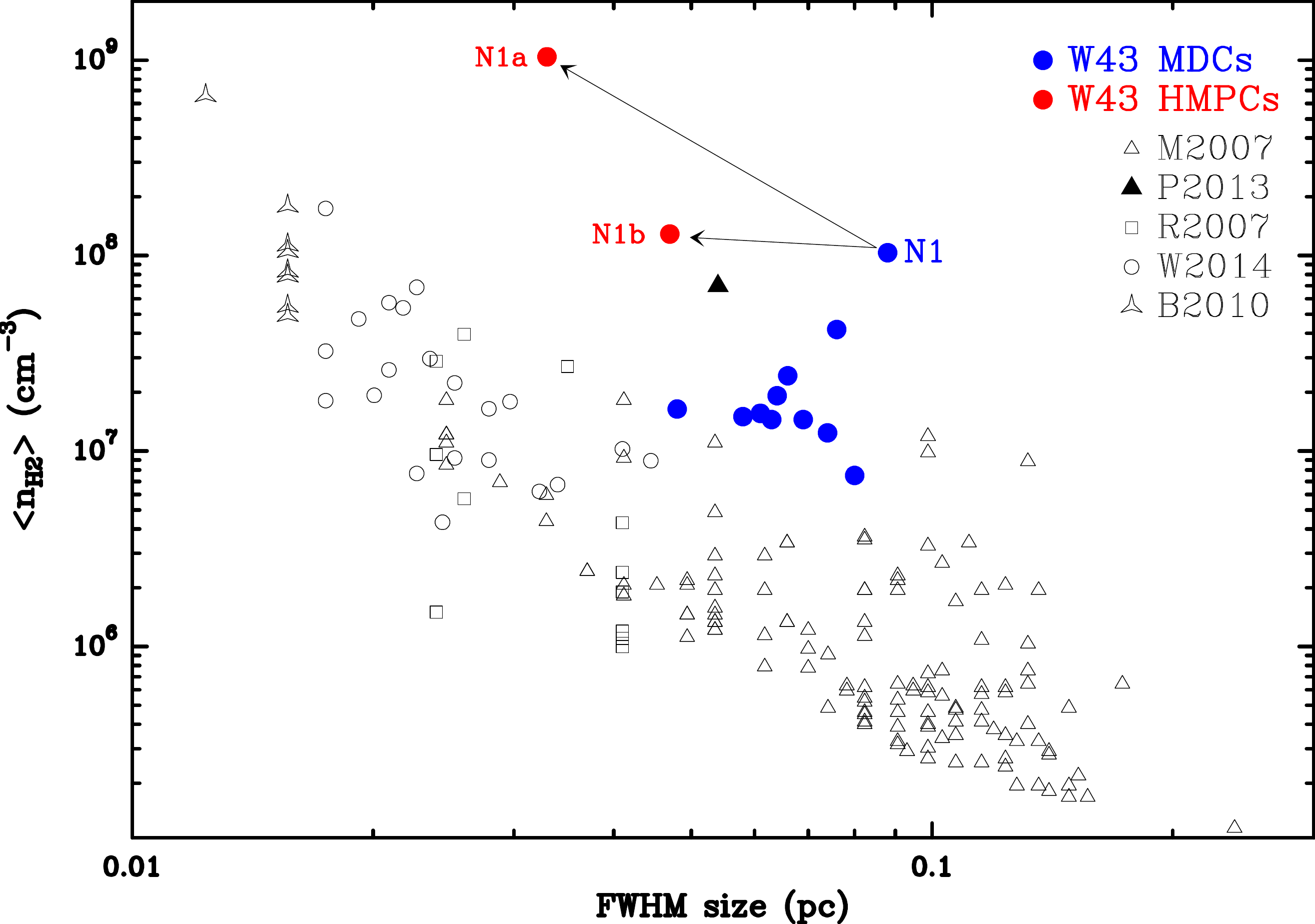}
 \caption{Comparison, density versus radius, of the W43 MDCs (massive dense cores) and HMPCs (high-mass protostellar cores) with the sources presented in \citet{motte07,rathborne07,bontemps10,peretto13,wang14}. The W43 MDCs and HMPCs lie above the general trend, with respect to their sizes, they are the densest cloud structures.}
 \label{f:dvsr}
 \end{center}
 \end{figure*}

Following the above assumptions on $\kappa$ and T=20 K, we have recalculated the published masses of a few MDC and protostar samples \citep{beuther02,rathborne07,motte07,bontemps10,peretto13,wang14} to make meaningful comparisons with present study (see Table~\ref{t:tableobj} and Fig.~\ref{f:dvsr}).

Because of obvious spatial resolution constraints, most search for high-mass protostars focused on $<3.5$~kpc regions. We recall that, at these close distances from the Sun, the richest high-mass star-forming region is Cygnus~X \citep[see][and  references therein]{kryukova14}.

W43-MM1 MDCs have radii which are twice smaller than those found in the MDCs of Cygnus~X \citep{motte07}, and they are ten times denser (see Table~\ref{t:tableobj}). As shown in Fig.~\ref{f:dvsr}, W43 MDCs lie above the general correlation of density versus radius found for samples of MDCs \citep{motte07} and HMPCs \cite[e.g.][]{rathborne07,peretto13,wang14}.

\subsection*{Special case of the remarkable objects N1a and N1b}

It is statistically understandable to find the most extreme objects in W43, since it is one of the most massive and most concentrated cloud complexes of the Milky Way \citep{quang11-w43}.

Nevertheless, we stress the remarkable N1a and N1b HMPCs extracted at 1 mm, which are $\sim$1100~$\msun$ and $\sim$400~$\msun$ respectively, gathering together 70\% of the N1 MDC mass measured at 3 mm. Their masses are consistent with those measured by \cite{sridharan14} when the difference of spatial resolution, dust mass opacity, and temperature are accounted for. These two HMPCs are a factor $30-90$ as massive as those found in Cygnus~X (see Table~\ref{t:tableobj} and Fig.~\ref{f:dvsr}). They look even more exceptional in comparison to high-mass protostars studied by \citet{wang14} (see Table~\ref{t:tableobj}).

N1a is even three times more massive and 15 times denser than the SDC335-MM1 HMPC \citep{peretto13} (see Table~\ref{t:tableobj}).
Remarkably, N1a is still in its earliest phase of evolution since it is only associated with weak mid-infrared emission \citep[][]{motte03}. Given its mass and following the definition of \citet{motte07}, N1a should host the most massive protostar known in the IR-quiet phase, i.e., before a $>$8~\msun~embryo has formed.

\section{Core Formation Efficiency of W43-MM1}
\label{s:cfe}


\begin{table*}[htbp!]
\begin{center}
\caption{Physical properties in subregions$^a$ of W43-MM1}
\label{t:sfr}
\addtolength{\tabcolsep}{-0pt}
\begin{tabular}{cccccc|ccc|ccc}
\hline
\hline
          &                &                    &                        &                                             &                                        & \multicolumn{3}{c}{\emph{Approach-1}}                                    & \multicolumn{3}{c}{\emph{Approach-2}}                \\ 
Cloud     & Area           &$M_{\rm cloud}~^b$  & $< n_{\rm H_2}>~^c$    & $M^{\rm \tiny total}_{\rm \tiny MDCs}~^d$   & CFE$^e$                                & $M_{\star}~^f$  & SFE$^g$                        & SFR$^h$               & $M_{\star}~^f$  & SFE$^g$     & SFR$^h$            \\
subregion & [pc$^2$]       &[\msun]             & [cm$^{-3}$]            & [\msun]                                     & [\%]                                   & [\msun]         & [\%]                           & [\msun\,Myr$^{-1}$]   & [\msun]         & [\%]        & [\msun\,Myr$^{-1}$] \\
          &(1)             &(2)                 &(3)                     &(4)                                          &(5)                                     &(6)              &(7)                             &(8)                    &(9)              &(10)         &(11)                \\
\hline                                                                                                                                                                     
A         & 1.06           & 8570               & 23.9$\times 10^4$      & 3055$^{~+190}_{-1630}$                      & $35.6^{~+2}_{-19}$                     & 915             & 10.7$^{~+0.6}_{-4.9}$          & 4575                  & 790             & 9.2         & 3950               \\ 
B         & 1.80           & 6900               & 5.0$\times 10^4$       & 645$\pm$155                                 & $9.3\pm2$                              & 195             & 2.8$\pm0.3$                    & 965                   & 375             & 5.4         & 1875               \\ 
C         & 2.90           & 4900               & 1.7$\times 10^4$       & 125$\pm$25                                  & $2.6\pm1$                              & 38              & 0.8$\pm0.3$                    & 190                   & 35              & 0.7         & 175                \\ 
D         & 6.62           & 3285               & 5.8$\times 10^4$       & $<$25                                       & $<1.5$                                 & $<$ 15          & $<$ 0.5                        & $<$ 75                & --              & --          & --                 \\ 
Ridge     & 5.75           & 20350              & 4.3$\times 10^4$       & 3825$^{~+370}_{-1530}$                      & 18.8$^{+2}_{-8}$                       & 1195            & 5.8                            & 5975                  & 1200            & 5.9         & 6000               \\
East      & 3.14           & 9955               & -                      & 190$\pm$45                                  & 1.9$\pm$0.5                            & 57              & 0.6                            & 285                   & --              & --          & --                 \\
West      & 2.74           & 10650              & -                      & 3635$^{~+325}_{-1820}$                      & 34$^{~+3}_{-17}$                       & 1090            & 10.2                           & 5450                  & --              & --          & --                 \\
\hline
\end{tabular}
\end{center}
~$^a$ The cloud subregions are defined in Sects.~\ref{s:cfe}-\ref{s:sfe-sfr} and Fig.~\ref{oncont}.\\
~$^b$ $M_{\rm cloud}$ is the mass derived by integrating the column density map built from \emph{Herschel} images.\\
~$^c$ The volumetric density is computed from Cols. 2 and 3 of Table~\ref{t:subregion} via $< n_{\rm H_2} >$= $M_{\rm cloud}/\rm{Volume}$.\\
~$^d$ Total mass of MDCs in the subregion. Uncertainty is the error on the extraction measurements (plus temperature for N1).\\
~$^e$ The core formation efficiency is computed from Cols. 2 and 4 via $CFE=M^{\rm \tiny total}_{\rm \tiny MDCs}/M_{\rm cloud}$.\\
~$^f$ Stellar mass derived by two approaches explained in Sect. \ref{s:sfe-sfr}.\\
~$^g$ The \lq instantaneous\rq~star formation efficiency is computed from Cols. 2 and 6 (resp. 9) via $SFE=M_{\star}/M_{\rm cloud}$.\\
~$^h$ The \lq instantaneous\rq~star formation rate estimated over a protostellar lifetime of $t_{\rm SF} = 0.2$~Myrs is computed from Col. 6 (resp. 9) via $SFR~=~M_\star/t_{\rm SF}$. \\
\end{table*}


\begin{figure*}[htbp!]
 \begin{center}
\includegraphics[trim=0cm 0cm 0cm 0cm, clip, scale=0.5]{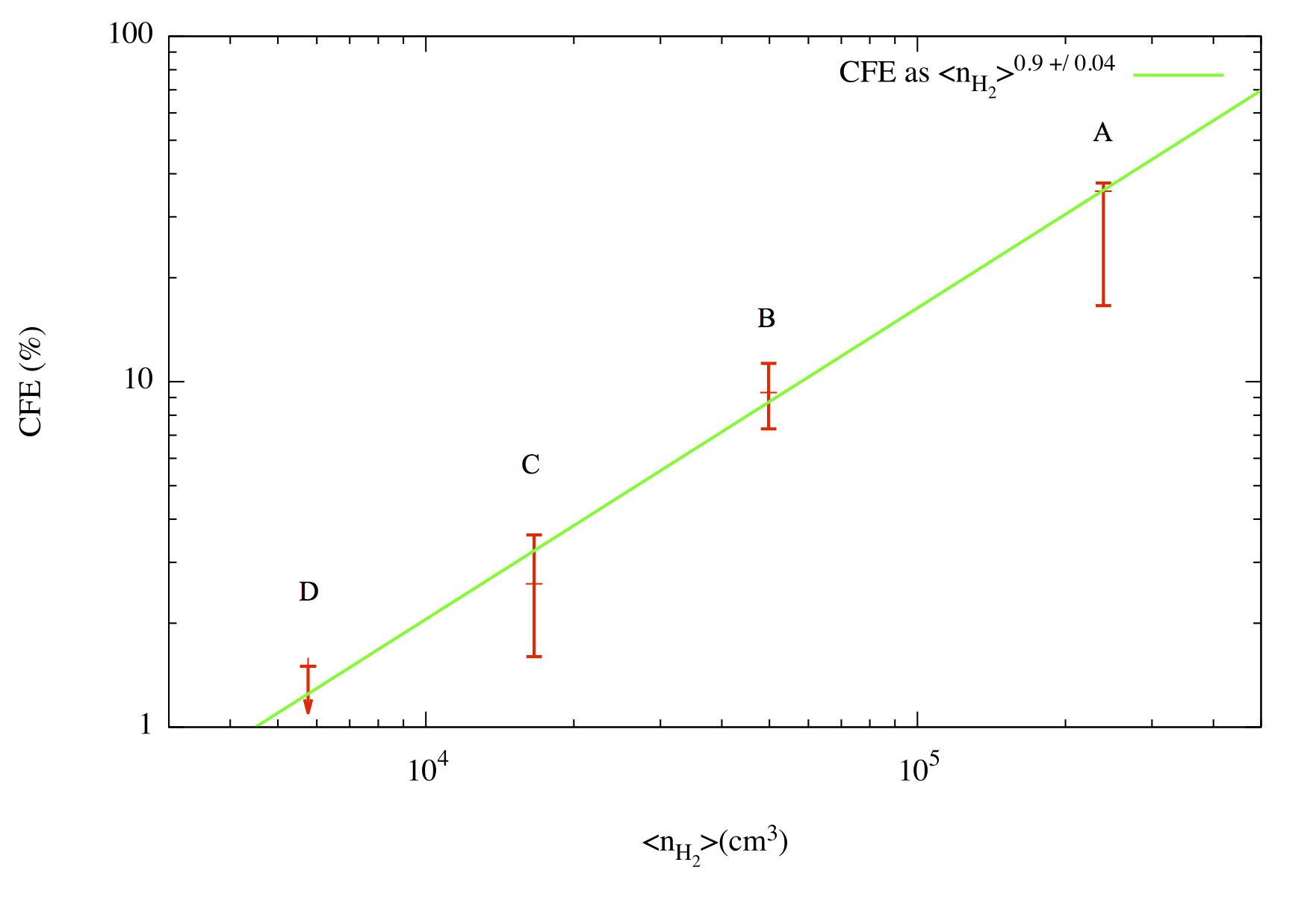}
 \caption{Linear correlation of the core formation efficiency from pc$^3$ clouds to $10^{-3}$~pc$^{3}$ dense cores with the cloud volume density. Relative uncertainties are estimated from error measures on MDCs masses and the green line is the power-law fit to these four points corresponding to subregions A-D.}
 \label{cfe-density}
 \end{center}
 \end{figure*}

We divided the ridge in four subregions A, B, C, and D (see Fig.~\ref{oncont}). Assuming\footnote{
The $\sigma\sim 2.2.$~km\,s$^{-1}$ turbulent velocity measured by \cite{quang13} cannot be used to estimate the crossing length since line widths in this region do not trace microturbulent motions but organized flows building the ridge (Louvet et al. in prep.).}  $c_{s}\simeq 0.2$~km\,s$^{-1}$, they are spaced from one another by more than five crossing lengths. This ensures that one generation of protostars takes place before MDCs change subregion. Translated in terms of column density, this leads to subregion A having \Nhtwo$~> 3.5\times10^{23}~\cmd$. Subregions B, C and D are then shells associated with the annular areas where  \Nhtwo~$\rm \in~[1.75-3.5]$, \Nhtwo~$\in~[1-1.75]\times 10^{23}~\cmd$ and \Nhtwo~$<10^{23}~\cmd$ respectively.
We hereafter define the CFE as the ability to concentrate pc$^3$ clouds with $n_{\rm H_2}\sim$ 10$^4~\cmc$ density into high-density seeds of $\sim$10$^{-3}$ pc$^3$ and $n_{\rm H_2}\sim$ 10$^7~\cmc$. This CFE connects those measured for 100~pc cloud complex to 1~pc clumps \citep[e.g][]{quang11-w48,eden12} to those for 0.1~pc MDCs to 0.01~pc protostars \citep[e.g.][]{motte98,bontemps10,palau13}.
We calculated the CFE as $CFE=M^{\rm \tiny total}_{\rm \tiny MDCs}/M_{\rm cloud}$, the ratio of the gas mass within MDCs over the subregion cloud mass.
 
The total mass of MDCs in each subregion, $M^{\rm \tiny total}_{\rm \tiny MDCs}$, was computed from the masses derived for MDCs extracted by \emph{Getsources}, as explained in Sect.~\ref{s:census} (see Table~\ref{t:tableobj}). The repartition of cores in the subregions A, B, C and D assumes that there are no projection effects, i.e., for instance, N9 belongs to B, not C or D. To check the robustness of this assumption we made tests randomly distributing MDCs in the different subregions. 
When N5, N7, and N11 MDCs, which should logically cluster in the high-density medium of subregion A, are located within subregion B, CFEs of A and B become 31\% and 15\% respectively. 
Thus, as long as the most massive MDCs N1 and N2 belong to subregion A, the slope index of the correlation discussed below between the CFE and the density changes by less than 20\%. 

\begin{figure*}[htbp!]
 \begin{center}
\includegraphics[trim=0cm 0cm 0cm 0cm, clip, scale=0.5]{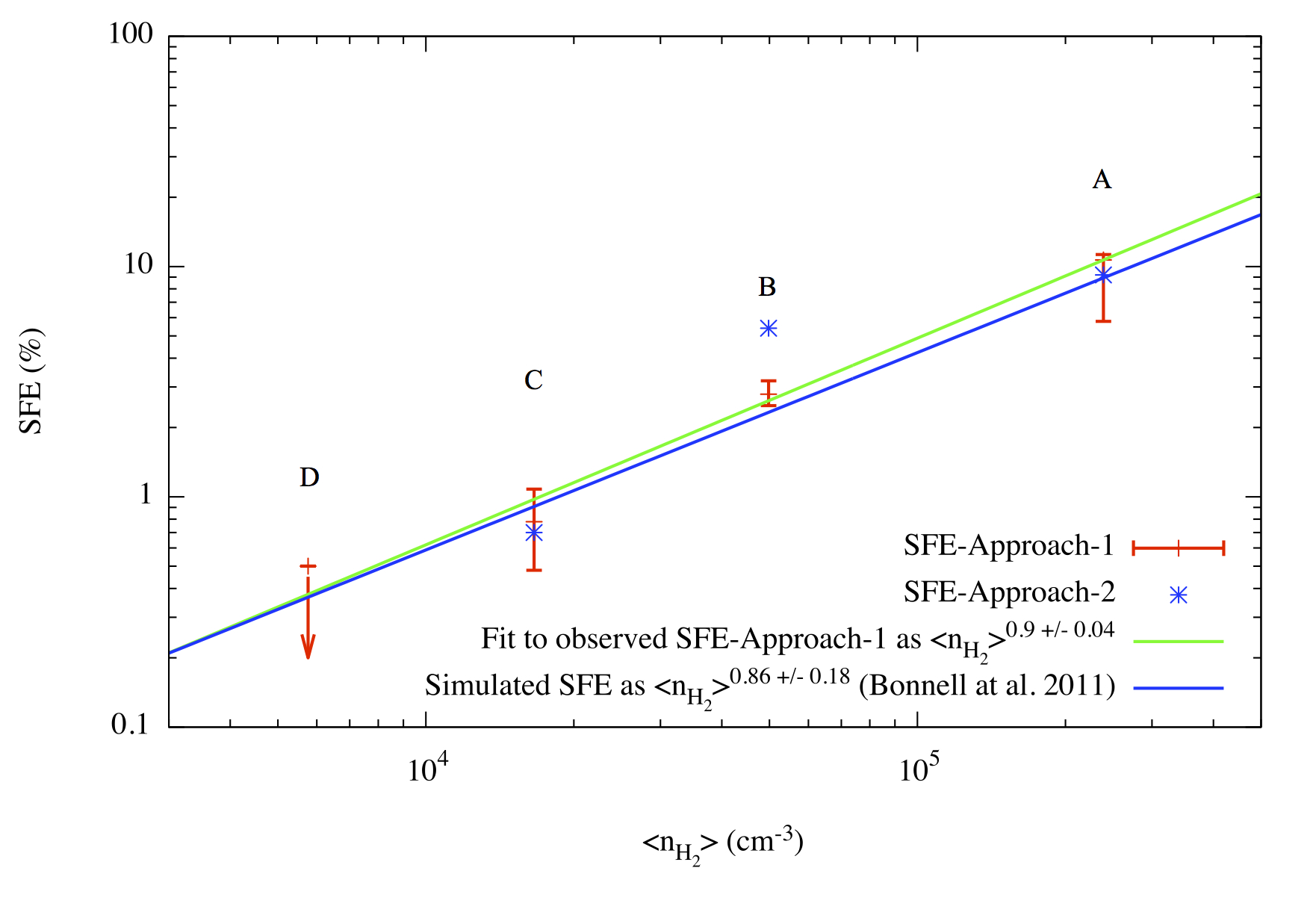}
 \caption{Linear dependence of the star formation efficiency in subregions A--D on cloud volume density. The red points and error bars correspond to the first approach explained in Sect. \ref{s:sfe-sfr} to derive the SFEs and blue points to the second. The green line is the linear fit to the first approach and the cyan line the relation of \cite{bonnell11}.}
 \label{sfe-dens}
 \end{center}
 \end{figure*}

The cloud masses, $M_{\rm cloud}$ (see Table~\ref{cfe-density}), were derived from the \emph{Herschel} column density map, after subtracting the $4\times 10^{22}~\cmd$ background level defined as in \citet{quang13}. To derive cloud densities, we had to define the 3D geometry of the subregions. Subregions are separated shells, in the sense that region B does not include region A, et cetera. Subregion A was taken to be a sphere of radius 0.53 pc. The subregions sums A+B, A+B+C, A+B+D+C are assumed to be ellipsoids with major axes in pc of  1.6 $\times$ 1.6 $\times$ 2.3, 2 $\times$ 2 $\times$ 3.9 and   2.1 $\times$ 2.1 $\times$ 4.6 respectively. For instance, the volume of region B, V$_B$, is then V$_{A+B}$-V$_A$=$\frac{4}{3}\times\pi\times\frac{1.6}{2}\times\frac{1.6}{2}\times\frac{2.3}{2}-\frac{4}{3}\times \pi\times$0.53$^3$ pc$^3$.
The relative uncertainties for volume densities of subregions A--D should be negligible and are not reported in Figs.~\ref{cfe-density}-\ref{sfe-dens}.
The absolute errors for cloud densities are $\sim$50\%, taking into account absolute uncertainties of $\sim$40\% for the cloud masses and 30\% error on cloud volumes due to line-of-sight effects.

Figure~\ref{cfe-density} displays the CFE measured for regions A--D as a function of their mean density. 
Relative uncertainties on the CFEs are $2-15\%$ only, since they only depend on the quality of the MDCs extraction. 
In contrast, absolute uncertainties could be as high as a factor of 4. This mostly comes from the combined inaccuracies of the dust mass emissivity both at \emph{Herschel} and 3.4 mm wavelengths. In Figs.~\ref{cfe-density}-\ref{sfrff}, we only consider relative uncertainties since we hereafter mainly discuss the relative behavior of the CFE (resp. the SFE) as a function of the cloud density.
Figure~\ref{cfe-density} reveals a clear correlation between the CFE and the cloud density, well represented by CFE~$\propto~<\nhtwo>_{\rm cloud}^{0.9}$. This slope has to be considered as a lower limit since the noise level of the 3.4~mm map increases towards its central part (i.e. from D to A), decreasing our MDC detection capabilities. 
With the relative CFE uncertainties and projection effects described above, the slope is uncertain by 5\% and 20\% respectively.

\cite{palau13} studied the fragmentation of a few $\sim$0.1~pc MDCs into $\sim$0.01~pc protostars. They gathered results from many other high-resolution millimeter studies and plotted the CFE against volume density. Their Fig.~6 displays the same trend as our W43-MM1 observations. The ability to concentrate gas seems thus to increase with density, for the cloud scale of 1-10~pc as well as the dense core scale of 0.1~pc, and possibly whatever the physical scale considered.


\section{Star formation efficiency and star formation rate}
\label{s:sfe-sfr}

\subsection{The instantaneous SFE and SFR}

We used the MDCs census from Sect.~\ref{s:census} to estimate the SFE and the SFR \citep[see also][]{motte03,quang11-w48,quang12}. The SFE is the ratio of the total mass of stars forming, $M_\star$, to the cloud mass $M_{\rm cloud}$ (see Table~\ref{t:sfr}) : $SFE=M_\star/M_{\rm cloud}$. The star formation rate itself is the ratio of stellar mass, $M_\star$, to the age of the star formation event considered. We have used a mean protostellar lifetime of $\sim$0.2~Myrs \citep{russeil10,ana13} to derive the SFRs. 

The above calculations give access to the \lq instantaneous' SFEs and SFRs. Indeed, the counting of protostars in each MDC should provide a direct measurement of star formation occurring in a cloud during one generation of protostars. It is especially adequate for ridges, which are forced-falling clouds \citep[see][]{schneider10,quang13}.
This contrasts with counts of \emph{Spitzer} young stellar objects (YSOs) in nearby clouds \citep[e.g.][]{heiderman10}, which compare the mass of already formed YSOs to the mass of a cloud forming a new generation of stars assuming a continuous star formation over 2~Myrs. These counts provide, by analogy, the `integrated' SFE \& SFR. 

\subsection{Calculation approaches}
The main difficulty encountered to estimate the instantaneous SFEs and SFRs is defining the total mass of forming stars, $M_\star$, in each dense core. We estimated $M_\star$, and thus the SFEs \& SFRs, using two approaches. The first one simply assumes that a constant efficiency from MDC to stellar cluster, $\epsilon$, is suitable for MDCs. The second one is based on an estimate of the most massive star each MDC can form, which is extrapolated to a protostellar cluster mass using the stellar initial mass function (IMF).

For the first approach in estimating $M_\star$, a `MDC to stellar cluster' efficiency of $\epsilon = 30 \%$ was assumed in the relation $M_\star=\epsilon\times M^{\rm \tiny total}_{\rm \tiny MDCs}$. At small scales, this core efficiency is generally assumed to be constant whatever the core mass \citep[e.g.][]{alves07}. The efficiency of $\epsilon = 30\%$ we use bridges the value estimated for the Cygnus X MDCs \citep[40\% in][]{bontemps10} and those measured for the lower-mass $\rho$~Oph dense cores \citep[$5-35\%$ in ][]{motte98}. It also reminds the efficiency measured by comparing the core mass function of low-mass star-forming regions to the IMF \citep[$\epsilon = 30\%$ according to][]{alves07, andre10}.
This approach directly relates the CFEs measured in Sect.~\ref{s:cfe} to the SFEs by \emph{SFE-Method1}~$= \epsilon\times$ \emph{CFE}. It is based on the assumption that $\epsilon$ does not depend on the MDC density, which is questionable according to e.g. \citet[][and references therein]{palau13}.

The second approach follows the finding of \citet{bontemps10} that Cygnus~X MDCs, which weigh $\sim$60~\msun~within 0.1 pc, form on average two ($\pm1$)  8~$\msun$ protostars. 
To be conservative, we assumed that MDCs less massive than $200~\msun$ would form only one $8~\msun$ protostar, plus its associated cluster. For cores N2 and N3, which are more massive than $200~\Msun$, given the result of \cite{peretto13} they should be able to form at least one  $50~\msun$ star. In the particular case of N1, the 1~mm data at 2$\farcs$2 show a fragmentation into two $\sim$0.04 pc cores, N1a and N1b. Each of them are above $200~\msun$, we therefore assumed that N1 would form two stars of 50 \Msun~plus their associated clusters. 

From these estimations of the most massive star forming in each MDC, we calculated the total stellar mass, $M_\star$ (see Table~\ref{t:sfr}), applying the canonical IMF description of \citet{kroupa2001}\footnote{ 
\citet{kroupa2001} describe the stellar IMF with a two-part power-law: $\xi(m)\propto m^{-\alpha_i}$ with $\alpha_i=1.3$ for $m \in[0.08,0.5]~\msun$  and  $\alpha_i=2.3$ for $m \in$ [0.5,$\infty$[~\msun.}. For this, we assumed that the IMF distribution applies to each subregion. It assumes that the detected MDCs along with the undetected lower-mass ones display a CMF that will correctly sample the IMF.
The IMF was integrated from the brown dwarf limit of $0.08~\msun$ to $150~\msun$ \citep{martins08,schnurr08},  leading to a fraction of stellar mass within high--mass ($>$ 8~\msun) stars of $\sim$22\%. The choice of 150 \msun~for the upper limit has not much impact on this stellar fraction, a choice of 300 \msun~\citep[see][]{crowther10} would lead to a stellar fraction within high--mass stars of $\sim$25\%.

Given the assumptions associated with these two approaches, the SFEs \& SFRs values we derived are consistent with each other. They agree  within factors of 1.15 to 2 (see Table~\ref{t:sfr}). 
The main limitation of the first method is the assumption that no star form outside the detected MDCs. As for the second method, the limitation comes from the applicability of the IMF to each subregions.
The relative and absolute uncertainties of the SFEs and the SFRs mostly arise from uncertainties on CFEs (see Sect.~4) and protostellar lifetime. 
We estimate relative uncertainties to be 20\% and 40\% and absolute ones to be 4 and 10 for the SFEs and the SFRs respectively.

\subsection{SFE relation with cloud density}

Figure~\ref{sfe-dens} displays, for the four subregions A, B, C, and D the SFE estimated through both approaches as a function of the cloud volume density. The correlation found between the CFE and density is retrieved for the SFE: $\rm{\emph{SFE-Approach1}}\propto<\nhtwo>_{\rm cloud}^{0.9\pm5\%}$. It obviously comes from the linear relation taken for the first approach but it validates the SFE values of the second approach, for which we fit \emph{SFE-Approach2}$\propto<\nhtwo>_{\rm cloud}^{0.9\pm22\%}$.


We then aim to compare the observed SFEs versus $<n_{\rm H_2}>_{\rm cloud}$ relation with those predicted by models. 
There is a lack of published plots that could be compared to our Fig.~\ref{sfe-dens}. We thus investigated the recent numerical simulations by \citet{bonnell11}, whose cloud mass and size as well as fragmentation resolution suit our present study. They simulated a 10$^4$ M$_\odot$, 10 pc elongated molecular cloud, initially globally marginally unbound due to turbulence, but with the high-mass star-forming region centered on the part of the cloud that is gravitationally bound.  Sink-particles are used to follow regions of gravitational
collapse over densities of 1.7$\times 10^{10}$ cm$^{-3}$  and sizes $<$~0.001 pc. We investigated the behavior of the SFE within shells around clumps against density and found a relation close to \emph{SFE}$~\propto ~<\nhtwo>_{\rm cloud}^{0.85}$ for $100-2\times 10^5$~\cmc~densities (see Fig.~\ref{ian} in Appendix).
 Astonishingly, this SFE relation is extremely close to the observed one and it clearly increases with density. Note that a similarly good correlation is found with volume density for the SFE measured within clumps rather than shells.
This behavior contrasts with past cloud--scale studies of the SFR \citep{evans09,lada10}. Indeed, they suggested a linear correlation, in log-log space, with the mass of the cloud above an $A_{\rm v}$ threshold, rather than its density (see however \citealt{gutermuth11}).

\subsection{SFE/SFR absolute values}

The SFEs obtained for the W43-MM1 ridge and its subregions A and B are large, SFEs$\,=3-11\%$ and their SFRs estimates are $4-11$ larger than the values predicted, given the subregion masses, by the simple equation proposed by  \cite{lada13}. Our estimations of the SFE and of the SFR over the ridge confirm its ability to form a rich cluster of massive stars: SFE = 6$\%$ and SFR = $6000~\msun$\,Myr$^{-1}$ over only a 8~pc$^3$ volume. These values are reminiscent of those found, on larger physical and time scales, for starburst galaxies \citep[see e.g.][]{kennicutt98}. As already noted by \citet{motte03}, the W43-MM1 ridge qualifies as a mini-starburst region. 
Like in the case of the G035.39--00.33 ridge \citep{quang11-w48}, both fragmentation and stellar formation are efficient in the high-density regions forming the W43-MM1 ridge.
The absolute value of the SFR of the W43-MM1 ridge has to be taken with caution due to the numerous uncertainties, but, it may account during one protostellar lifetime of $\sim$0.2~Myr for one twentieth of the total $\sim$1~\msun\,yr$^{-1}$ SFR of the Milky Way.

A closer inspection of star formation activity between the eastern and the western parts of the ridge reveals a clear disparity (see Fig.~\ref{oncont} and Table~\ref{t:sfr}).
Despite the fact that they have similar masses, the SFE and the SFR in the western part are about twenty times larger than in the eastern one (10.2\% versus 0.6\%).
With a microturbulent support alone, the cloud of the eastern ridge would instantly fragment and form stars. However, the W43-MM1 ridge is constituted of several gas flows/filaments that are supersonically merging (Louvet et al. in prep.) and developing shears and low-velocity shocks \citep{quang13}. 
The observed SFE disparity is thus coherent with the western part of the ridge having already formed a protostellar cluster (see Fig.~\ref{oncont}) and its eastern part still being assembling material (Louvet et al. in prep.).

\section{Comparison to statistical models of star formation rate}
\label{s:sfrff}


\begin{table*}[htbp!]
\begin{center}
\caption{SFR$_{\rm ff}$ in W43-MM1}
\label{t:subregion}
\addtolength{\tabcolsep}{-2pt}
\footnotesize{
\begin{tabular}{cccccc|cc|ccc}
\hline
\hline
Cloud &     $\alpha_{vir}$         & Volume$^a$      & $M_{\rm cloud}^b$  & $<\nhtwo>_{\rm cloud}^c$ & $t_{\rm\tiny ff~ cloud}~^d$   & $<\nhtwo>_{\rm cores}^e$   & $t_{\rm\tiny ff~ MDCs}~^f$      & SFR$_{\rm ff}~^g$       & SFR$_{\rm ff}~^g$         \\
region      &[-]                   & [pc$^3$]        & [\msun]            & [\cmc]                   & [kyrs]                        & [10$^7$\cmc]               & [kyrs]                          & (\emph{Approach-1}$^h$) & (\emph{Approach-2}$^h$)   \\
            &(1)                   &(2)              &(3)                 &(4)                       &(5)                            &(6)                         &(7)                              & (8)                     & (9)                       \\
\hline                                                                                                                                                                                                                                                       
A           & $0.11\pm0.07$        & $0.6$           & 8570               & $2.39\times 10^5$        & 70                            & $4.7^{+0.3}_{-2}$          & 4.9$^{+0.2}_{-1.2}$             & $1.50^{+0.15}_{-1.1}$   & 1.30                      \\ 
B           & $0.22\pm0.13$        & $2.4$           & 6900               & $4.98\times 10^4$        & 150                           & $1.7\pm0.5$                & 8.2$\pm1.4$                     & $0.50\pm0.22$           & 1.00                      \\ 
C           & $0.40\pm0.24$        & $5.1$           & 4900               & $1.66\times 10^4$        & 260                           & $0.9\pm0.2$                & 11.2$\pm$1.2                    & $0.18\pm0.06$           & 0.16                      \\ 
D           & $0.74\pm0.45$        & $9.9$           & 3285               & $5.76\times 10^3$        & 445                           & $<$0.9                     & $>$11.2                         & $<$0.09                 & -                         \\ 
\hline
\end{tabular}}
\end{center}
~$^a$ Volumes derived from assumptions exposed in Sect.~\ref{s:cfe}.\\
~$^b$ $M_{cloud}$ is the mass derived by integrating the column density map built from \emph{Herschel} images.\\
~$^c$  The volume density of the cloud is computed from Cols. 2 and 3 via $<n_{H_2}> = {M}_{\rm cloud}/\rm{Volume}$. \\
~$^d$ The freefall time of the cloud is computed from the cloud density (Col. 4) and Eq. 3.\\
~$^e$ The volume density of the MDCs is the ratio of the total mass of the cores, M$^{\rm total}_{\rm MDCs}$, to the sum of MDCs' volume (see Table~3). \\
~$^f$ The freefall time of the MDCs is computed from the MDCs density (Col. 6) and Eq. 3.\\
~$^g$ $SFR_{\rm ff}$~is estimated via Eq.~\ref{e:pat}, Cols. 5 and 7, and SFE values of Table~3.\\
~$^h$ The \emph{Approach-1} (resp. \emph{Approach-2}) refers to the first (resp. second) approach in estimating the SFE presented in Sect.~\ref{s:sfe-sfr}.
\end{table*}


\begin{figure*}[htbp!]
 \begin{center}
\includegraphics[trim=0cm 0cm 0cm 0cm, clip, scale=0.5]{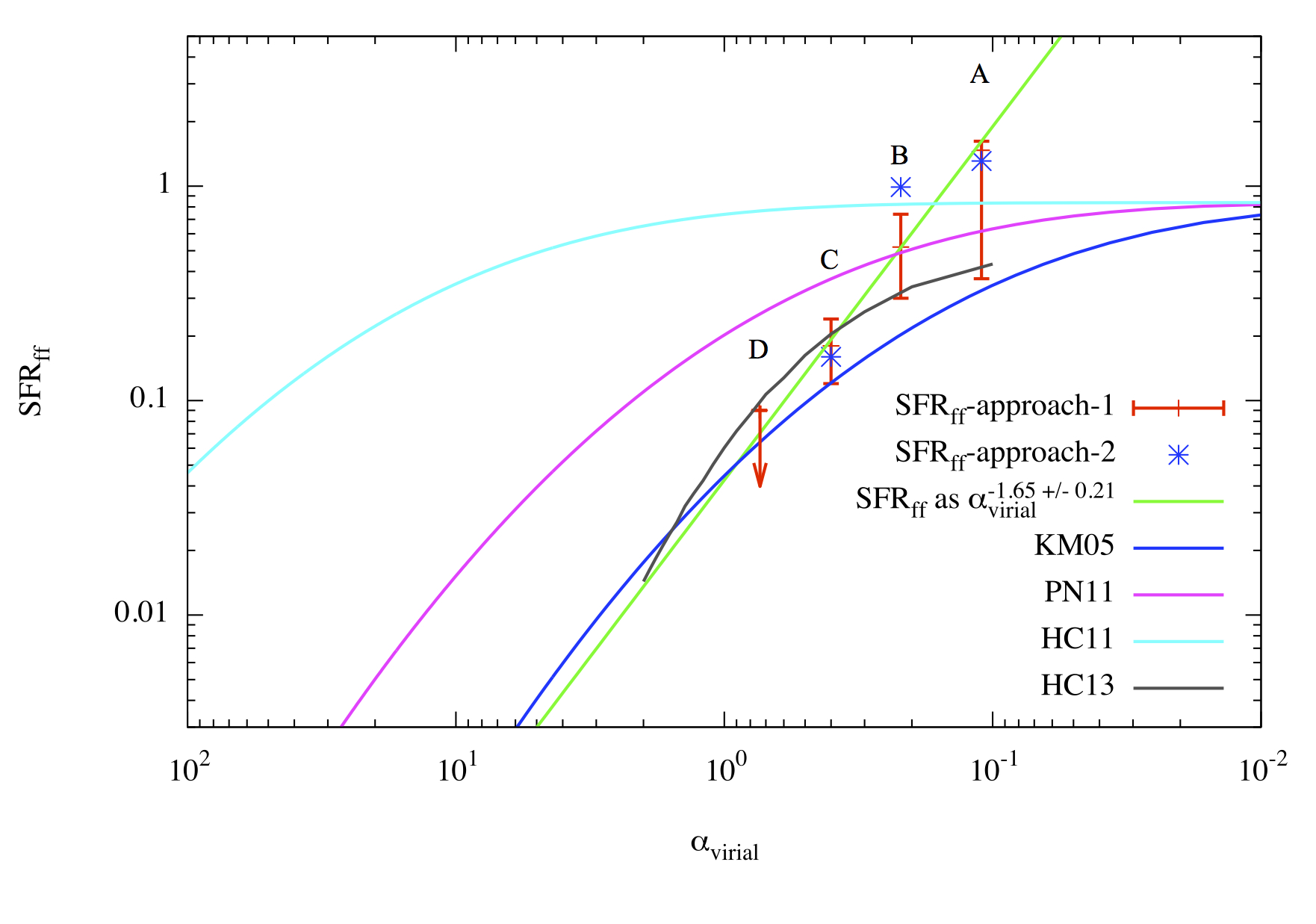}
 \caption{ SFR estimates over W43-MM1 compared to the multi-freefall extrapolation of models from \citet{krumholz05}, \citet{padoan11}, \citet{hennebelle11}, given in \citet{federrath12} and a magnetized model from \citet{hennebelle13} (blue, pink, cyan and black curves respectively). The analytic models struggle to reproduce the observed SFR$_{\rm ff}$ at low $\alpha_{vir}$.}
 \label{sfrff}
 \end{center}
 \end{figure*}


The SFR statistical models are analytic descriptions of a turbulent cloud including magnetized turbulence and self-gravity, which acts as a filter to select the 
core progenitors \citep{krumholz05,padoan11,hennebelle13}. They are based on the integration of the density probability distribution function (PDF), 
which is assumed to be a log-normal distribution. In the simplest approach, 
the density PDF is weighted by the free-fall time and integrated above a certain density threshold.  
Therefore, the different models differ by the density threshold which is chosen in the integration of the PDF \citep[see e.g.][]{hennebelle11, federrath12}. 
However in practice, while reasonable, this approach does not take into account the complex and heterogeneous spatial distribution of the gas. Moreover, 
it is rather unclear that simple thresholds, based for example on mean Jeans mass, are justified since the 
density varies over orders of magnitude.
A different type of approach is the calculations performed by  \cite{hennebelle11,hennebelle13} 
\citep[see also multi-freefall extrapolations of SFR models by][]{federrath12}.
They use the multi-scale method developed in cosmology \citep{press74} and take into account the gas spatial distribution characterized by a power spectrum. If the density variance is scale independent, it is equivalent to the PDF integration approach described above.
Due to our lack of knowledge in the exact statistics of molecular clouds, in particular how to define their boundaries, 
this approach is also hampered by large uncertainties.

To infer a dimensionless star formation rate  independent of the cloud mass and density, SFR$_{\rm ff}$ \citep{krumholz05}, the following normalized quantity has been defined:
\begin{align}
\rm\emph{SFR}_{\rm ff}&=\frac{M_\star}{M_{cloud}}\times\frac{t~_{\rm ff}^{\rm cloud}}{t~_{\rm ff}^{\rm MDCs}} \nonumber\\
&=\emph{SFE}\times\frac{t~_{\rm ff}^{\rm cloud}}{t~_{\rm ff} ^{\rm MDCs}}~.
\label{e:pat}
\end{align}

The freefall times, $t~_{\rm ff}^{\rm cloud}$ and  $t~_{\rm ff}^{\rm MDCs}$, are estimated from the mean cloud and mean MDCs densities respectively via:
\begin{equation}
t_{\rm ff}(\rho)\equiv\left(\frac{3\pi}{32~G~\rho}\right)^{1/2}
\label{e:tff}
\end{equation}
where G is the gravitational constant and $\rho$ the cloud or MDCs gas density.

Usually, models plot the SFR$_{\rm ff}$ as a function of the virial parameter, $\alpha_{\rm vir}= 2E_{\rm kin}/|E_{\rm grav}|
=\sigma^2\times R_{\rm cloud}\times 5/(3\times G\times M_{\rm cloud})$ because they are two normalized quantities.
We thus have computed the freefall times of all subregions A, B, C, and D plus the mean freefall time of the MDCs they host (see Table~\ref{t:subregion}). 
We have estimated $\alpha_{\rm vir}$ for all subregions\footnote{The radii of subregions B, C, and D are estimated from spheres with volumes equal to V$_B$, V$_C$ and V$_D$ respectively.} (see Fig.~\ref{oncont} and Table~\ref{t:subregion}), using a turbulence velocity of 
$\sigma$ = 2.2~km s$^{-1}$ \citep{quang13} adequate for the complete W43-MM1 ridge. We calculated the SFR$_{\rm ff}$ for the two approaches (see Table~\ref{t:subregion}) presented in Sect.~\ref{s:sfe-sfr}.
We selected models of Mach number equal to 9.5 and parameters proposed in \cite{federrath12}\footnote{The forcing parameter, $b$, is set to 0.4; the magnetic field is not taken into account ($\beta\rightarrow\infty$).}. 
As described above, these three models integrate the density PDF and differ by the density thresholds. The model 
labeled KM05 uses the sonic length, i.e. the length at which velocity dispersion and sound speed are equal, and requires that the Jeans length be smaller that its value. The 
model labeled PN11 requires that the Jeans length must be smaller than the typical size of the shocked layer,
 while the model labeled HC11 simply states that the integration should be performed over all pieces of gas 
whose densities are such that the associated Jeans length is smaller than a fraction of the cloud size. 
 Second, since a significant magnetic field has been measured toward W43-MM1 (mass-to-flux ratio $\sim$2,~\citealt{cortes10}), we also present a magnetic model
taken from \citet{hennebelle13}. For this model a magnetic field of $20~\mu$G$\times(\nhtwo / 10^3~\cmc)^{0.3}$ is assumed. 
These values are reasonable given what is known on the magnetic field in this region \citep{cortes10}. Their exact choice, at this stage, is dictated by the reasonable agreement with the data on the SFR. 

Before comparing our results with the models, we would like to stress that all models have similar behaviors. Indeed, for extremely cold clouds (i.e. have low $\alpha$), most of the gas is gravitationally unstable and therefore a significant fraction of the density PDF contributes in the integration. Since it is normalized by the total mass and mean freefall time, the SFR tends toward a constant value which is of the order of $\epsilon$. 
However, because the density PDF is weighted by the freefall time which is shorter at high densities, the normalized SFR can be larger than $\epsilon$, 
here taken to be 30\%. For clouds which are more supported against gravity, only the densest regions contribute to star formation. Thus only the high-density part of the PDF contributes. This leads to a SFR which can be arbitrarily low and can have a stiff dependence on the cloud parameters. 

Figure~\ref{sfrff} displays our SFR$_{\rm ff}$ estimates against $\alpha_{\rm vir}$ along with the multi-freefall extrapolation of isothermal models 
from \cite{krumholz05} and \cite{padoan11} computed by \cite{federrath12}, plus the magnetized model exposed in \cite{hennebelle13}. 
 Like in models, the observed SFR$_{\rm ff}$ relation at high $\alpha_{\rm vir}$ increases with decreasing virial parameter $\alpha_{\rm vir}$ and its dependence index recalls the one of models at high $\alpha_{\rm vir}$ (see Fig.~\ref{sfrff}). But, none of the models can correctly describe the observations at low $\alpha_{\rm vir}$ ($<0.2$). Indeed, all models depart from the $\alpha_{\rm vir}$-dependent regime to join the saturation regime while observations seem still to be anti-correlated to $\alpha_{\rm vir}$. We nevertheless note the HC13 model is in better agreement and that the second approach in estimating SFR$_{ff}$ has a trend closer to the model behavior.

Both theories and observations need to go one step farther to solve this question. For theories, the difference in behavior could be understood when recalling that ridges do not fit two major hypotheses of these analytic models. Ridges first represent column density points that depart from the log-normal distribution assumed in all models \citep{hill11-vela}. Second, these regions are forced-falling clouds whose turbulence level probably does not follow the Larson law used in the SFR$_{\rm ff}$ models \citep{schneider10,quang13}. Combined, these two reasons could explain why in gravity-dominated regions the current SFR$_{\rm ff}$ models cannot apply in their  present formulation. From the observational side, a higher resolution and deeper imaging are necessary to estimate robust SFR values from a complete census of high- to low-mass protostellar cores.

\section{Conclusion}
\label{s:conclusion}

We use the IRAM Plateau de Bure interferometer to image the W43-MM1 ridge at 3~mm and a zoom on its main MDC at 1~mm (see Fig.~\ref{oncont}). We compare the mass distribution observed throughout these maps with the column density image of W43-MM1 built from \emph{Herschel} data. Our main results and conclusions may be summarized as follows:
\begin{itemize}
\item The 3~mm mosaic reveals eleven $\sim$0.07~pc MDCs, labeled N1 to N11, across the W43-MM1 ridge. These MDCs range in mass between $\sim$50~$\msun$ and $\sim$2100~$\msun$; have mean densities between $n_{\rm H_2}$ $\sim 7\times10^6~\cmc$ and $\sim$1$\times10^8~\cmc$. The 1~mm snapshot identifies two $\sim$0.03--0.04~pc HMPCs within N1, the most massive of the MDCs sample (see Table~1).
The N1a protostellar core, with its $\sim$1080~$\msun$ mass, is the most massive known 0.03~pc young stellar object ever observed in an early phase of evolution (see Fig. 2). It is expected to form a couple of $\sim$50~$\msun$ stars.
\item We use the MDCs masses to estimate the concentration of the cloud gas toward high density (see Table 3), usually called the gas-to-core formation efficiency (CFE). The W43-MM1 ridge split into four exclusive subregions displays a clear correlation of the CFE with cloud volume density: \emph{CFE}~$\propto~<\nhtwo>_{\rm cloud}^{0.91}$  (see Fig.~\ref{cfe-density}).
\item The CFE measurements are extrapolated to \lq instantaneous' stellar formation efficiencies (SFEs) following two approaches constraining the MDC to stellar cluster efficiency (see Table 3 and Fig. 4). The SFE values are also used to make estimate of 1) the `instantaneous' stellar formation rate (SFR) expected during the protostellar lifetime and 2) the dimensionless star formation rate per free-fall time theoreticians use: SFR$_{\rm ff}$.
\item The SFEs obtained for the W43-MM1 ridge and its subregions A and B are large, SFEs$\,=3-11\%$, and their SFRs estimates are $4-11$ times larger than the values expected from their masses, following the equation proposed by \cite{lada13}. We propose it is due to a strong correlation of the CFE to the gas volume densities in W43-MM1.
With its SFR absolute value, \emph{SFR}~$ = 6000~\msun$\,Myr$^{-1}$, W43-MM1 qualifies as a mini-starburst region. It may account, during one protostellar lifetime of $\sim$0.2~Myr, for as much as one twentieth of the total $\sim$1~\msun\,yr$^{-1}$ SFR of the Milky Way.
\item The CFE of the eastern and western parts of the ridge are clearly unbalanced, leading to SFE values as different as 0.6\% and 10.2\%. It might be due to the eastern region currently assembling its mass along multiple filaments whose interaction could impede cloud fragmentation and star formation.
\item Our observations lead to a SFR$_{\rm ff}$ relation with virial number which is steadily increasing when $\alpha_{\rm vir}$ is decreasing (see Fig. 5). While statistical SFR models display such a trend for high $\alpha_{\rm vir}$, they saturate for values close to those observed in the W43-MM1 ridge. Models with more realistic conditions are necessary to fully describe the complexity of this very dense, turbulent, non isothermal, and non stationary cloud structure. Higher resolution and deeper imaging are necessary to confirm current observational findings.
\end{itemize}

\begin{acknowledgements}
We thank Christoph Federrath for the fruitful discussions we had on SFR. We are grateful to Alexander Men'shchikov for help in customizing \emph{Getsources} for interferometric images.
\end{acknowledgements}

\bibliographystyle{aa}
\bibliography{fab}

\newpage

\section{Appendix}
\label{app}

\begin{figure*}[htbp!]
\begin{center}
\includegraphics[trim=0cm 0cm 0cm 0cm, clip, scale=0.5]{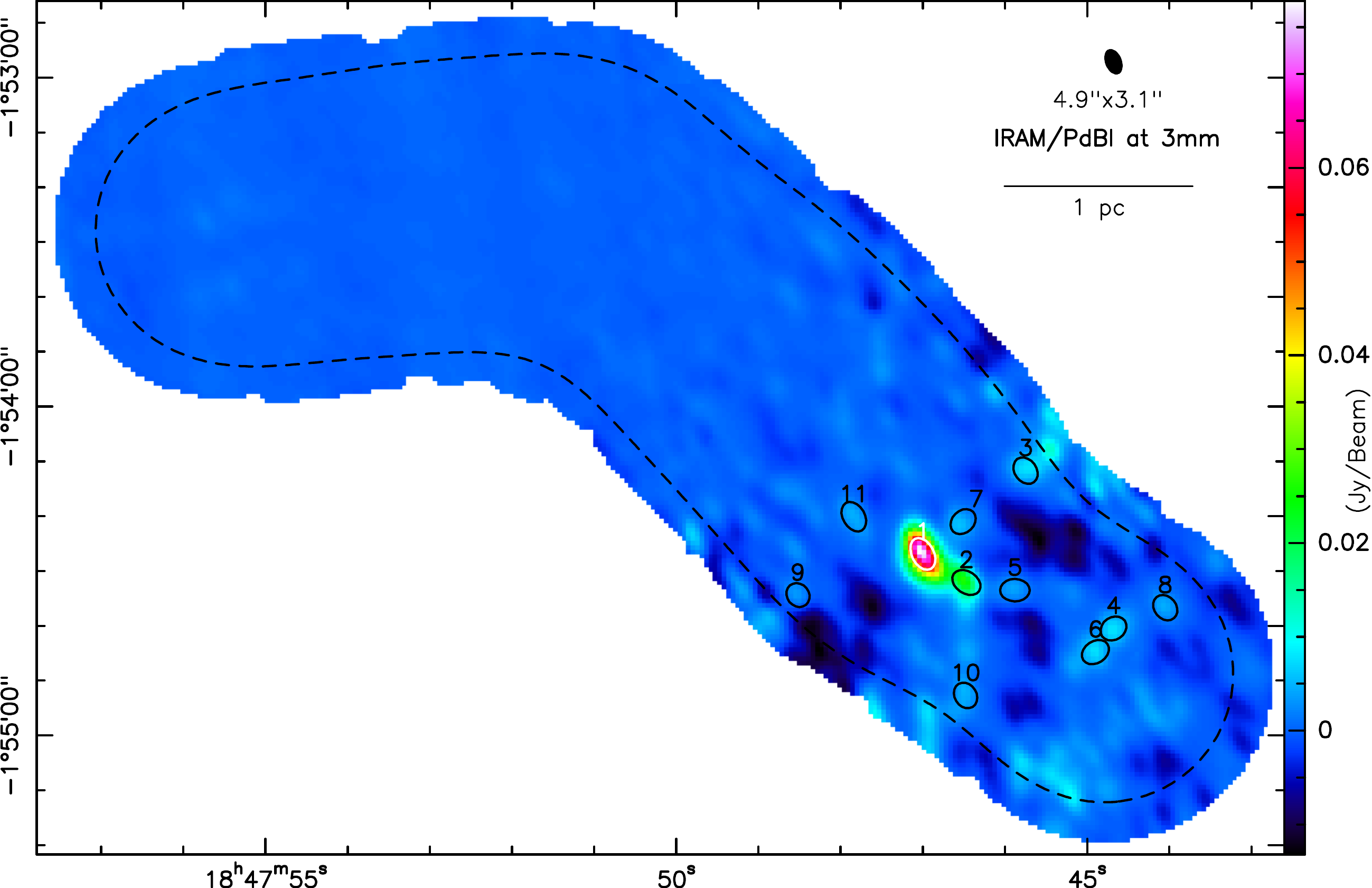}
\caption{IRAM/PdBI 3~mm continuum image of the W43-MM1 ridge the black dashed contour delimit the area where the confidence map exceeds 10\%. Black and white ellipses plus numbers locate MDCs extracted by \emph{Getsources}.}
\label{fig1-ap}
\end{center}
\end{figure*}

\begin{figure*}[htbp!]
\begin{center}
\includegraphics[trim=0cm 6cm 1cm 1cm, clip, scale=0.5]{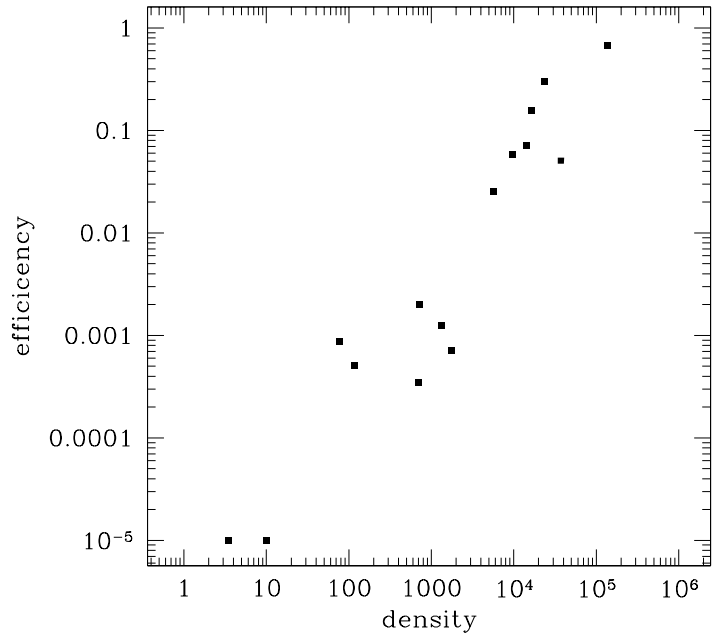}
\caption{Efficiency of core formation measured as the
fraction of mass that is inside the SPH sink-particles as a function of the cloud gas density (extrapolated from \citealt{bonnell11}).
Both the efficiency and the gas densities are measured in spherical shells centered on the
densest region of the simulation and span size-scales from $\sim0.04$ to 10 pc. The points
represent regions where the sink-particles are all between 40,000 and 100,000 years old. The efficiencies
increase with time such that older systems would have higher efficiencies, but the relation between CFE and $<n_{H_2}>$ remains.}
\label{ian}
\end{center}
\end{figure*}

\end{document}